# Sesame-Style Decomposition of KS-DFT Molecular Dynamics for Direct Interrogation of Nuclear Models

Sarah C. Burnett[1,2 a)], Daniel G. Sheppard[1, b)], Kevin G. Honnell[1, c)], and Travis Sjostrom[1, d)]

[1]*Los Alamos National Laboratory, New Mexico 87545*
[2]*University of Maryland, College Park, MD 20742*

a)burnetts@math.umd.edu
b)danielsheppard@lanl.gov
c)kgh@lanl.gov
d)sjostrom@lanl.gov

**Abstract.** A common paradigm used in the construction of equations of state is to decompose the thermodynamics into a superposition of three terms: a static-lattice cold curve, a contribution from the thermal motion of the nuclei, and a contribution from the thermal excitation of the electrons. While statistical mechanical models for crystals provide tractable framework for the nuclear contribution in the solid phase, much less is understood about the nuclear contribution above the melt temperature ($C_v^{(nuc)} \approx 3R$) and how it should transition to the high-temperature limit ($C_v^{(nuc)} \sim \frac{3}{2}R$). In this work, we describe an algorithm for extracting both the thermal nuclear and thermal electronic contributions from quantum molecular dynamics (QMD). We then use the VASP QMD package to probe thermal nuclear behavior of liquid aluminum at normal density to compare the results to semi-empirical models -- the Johnson generic model, the Chisolm high-temperature liquid model, and the CRIS model.

## INTRODUCTION

For the past sixty years, a common paradigm for constructing broad-ranging equations of state (EOS) has been to decompose the overall thermodynamics into a superposition of three terms: the static-lattice cold curve contribution, $\phi_0(\rho)$, the thermal nuclear contribution, and the thermal electronic contribution [1-4].

$$F(\rho, T) = \phi_0(\rho) + F_{nuc}(\rho, T) + F_{el}(\rho, T) \qquad (1)$$

where $F$ denotes the Helmholtz free energy; $\rho$, the density; and $T$, the temperature. For example, the Sesame EOS library, developed at Los Alamos [5], makes extensive use of this decomposition in its constituent EOS tables. Other thermodynamic properties of interest (e.g., pressure, internal energy, heat capacity) can then be obtained from the appropriate derivatives of $F$ [2]. Though Eq. 1 offers a tractable framework for modeling EOSs, it implicitly assumes that the three contributions are uncoupled – i.e., that the contributions from motion of the nuclei are independent from and uninfluenced by the degree of electronic excitation, and vice-versa. This approximation cannot be assessed or validated by experiment and, thus, remains an unquantified source of uncertainty in the resulting EOSs.

In this paper, we describe a tractable way for assessing the approximations inherent in Eq. 1, extracting the thermal nuclear and thermal electronic contributions via quantum molecular dynamics (QMD) simulations. We illustrate the methodology using the Vienna *ab initio* Software Package (VASP) [6-9] to examine liquid aluminum at 2.7 g/cm$^3$ and temperatures between 2,000K and 47,000K. Results for the constant-volume heat capacity are compared to empirical models for the thermal nuclear term: the Johnson "generic" model [10], the Chisolm "high-temperature" model [11], and the Kerley CRIS model [12-14].

Our current understanding of the EOS is strong in two regimes: the solid state which can be modelled as phonons vibrating as seen in the Debye [15] or Einstein [16] models and the ideal gas limit at high temperatures where atoms are colliding independently. In the crystalline solid as the temperature is increased, the heat capacity approaches 3R at the Dulong-Petit limit [17] near the melting temperature (where R denotes the gas constant). Continuing along an isochore, at very high temperatures the nuclear (ionic) component of the heat capacity eventually decays to $\frac{3}{2}R$, the heat capacity of a monatomic ideal gas. For modelling purposes, above melt, the Debye/Einstein contribution is switched off as a function of temperature, resulting in only the contribution from kinetic energy of atoms at high temperatures. It is our goal to examine the intermediate liquid/fluid regime, between the melting point and the high temperature asymptote, as exhibited in Fig. 1. In this work, we present quantum molecular dynamics (QMD) simulations of the nuclear component that would otherwise not be obtainable in an experimental setting where the heat capacity cannot be decomposed into separate contributions from nuclear motion and electronic excitation.

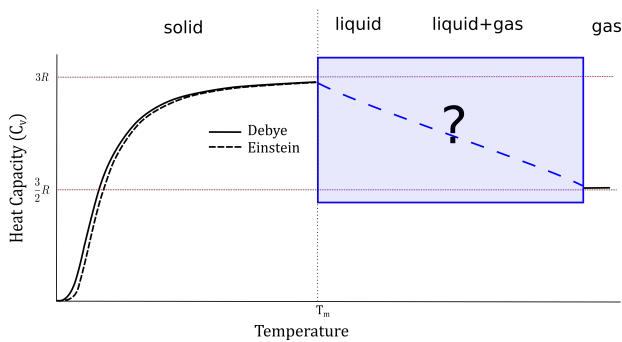

**FIGURE 1.** The heat capacity is well known in the solid and gas states but the liquid state compels further investigation.

## DENSITY FUNCTIONAL THEORY DETAILS

First-principle methods are necessary to understand the behavior on the atomic scale in its decomposed form. QMD uses Kohn-Sham (KS) density functional theory (DFT) [18] to obtain the electronic structure in the Born-Oppenheimer approximation and propagates the nuclei classically in time. The computational tool known as the Vienna *ab initio* Simulation Package (VASP) [6-9] eases some of the computational work with its built-in software for atomic-scale modeling. The plane-wave pseudopotential code implements a generalized gradient approximation (GGA), specifically, the Perdew-Burke-Ernzerhof (PBE) [19,20] exchange-correlation functional and utilizes a three-electron PAW potential. The orbitals are filled via Fermi-Dirac distribution consistent with the temperature of the simulation; we refer to this temperature as $T_{el}$. All the liquid calculations are performed at the gamma-point as long-range symmetry constraints are not desired for the liquid phases. A plane-wave energy cut off of 450 eV was determined to be sufficient.

Simulations of elemental aluminum were run with 100 atoms and a constant density of 2.70 g/cm$^3$. Studies revealed that 50 atoms produced different results (due to the periodic finite size effect) when compared to 100 atoms with a >5% relative error in their values for energy per atom. The relative error between 200 atoms and 100 atoms was measured at <1% so 100 atoms was chosen as the optimal choice. These studies were aimed to investigate the liquid regime at different temperatures, along an isochore, so the simulations were run every 1,000K from temperatures of 2,000-47,000 K (0.1724-4.052 eV). Trials too close to the melting temperature of 933.47K [21] exhibited evidence of solidification. Simulations became more expensive as the temperatures increase due to the necessity of including more Kohn-Sham orbitals in the high temperature runs. As an example, our T=10,000K calculation required ~432 bands whereas T=45,000K required ~1931 bands. We achieve a threshold occupation of $10^{-5}$ for all temperatures.

Trials were run for 20 ps, with the initial 2 ps removed so that we used data that had equilibrated, and then extracted the energy by averaging. A time step of 0.5 fs was implemented for trials up to T=5,000K and 0.2 fs for higher temperatures up to T=70,000K. Corresponding microcanonical ensemble simulations were executed to ensure these time steps conserved energy. The energies extracted contained all components of the decompositions. In order to remove the thermal electronic contribution, we sampled the data by extracting the position of the atoms every 5 fs after removing the first 2 ps of data. A low-temperature electronic structure calculation was performed on each of the sampled high-temperature configurations by setting the number of steps in the sample runs to 0 and the electronic temperatures to 0.001 eV. (For numerical robustness, a small, finite electron temperature is employed, and then extrapolated to zero, to ensure convergence stability [22]). The total energy was divided by 100 (the number of atoms) and a factor of $\frac{3}{2}RT$ was added as the kinetic energy contribution. The energies so obtained retain just the cold curve contribution and the thermal nuclear contribution. Since the cold curve contribution only depends on the density, by taking the derivative of these energies with respect to temperature, we obtain a heat capacity at constant density that contains just the thermal nuclear component. Finally, this data was ready to compare to the semi-empirical models.

In summary, our algorithm is:

- Run standard QMD at fixed density and temperature, where $T_{nuc} = T_{el}$.
- Allow configuration to equilibrate and discard equilibrium initialization period.
- Sample and store configurations from equilibrated trajectory at an interval greater than the correlation length, to ensure a non-correlated sampling.
- Run single point calculations through the sampled configurations at 'zero' electronic temperature.
- Average the energy from the single-point calculations, divide by the total number of atoms, and add the thermal kinetic energy ($\frac{3}{2}RT$).
- Numerically differentiate energy with respect to temperature to get the thermal nuclear heat capacity.

## RESULTS

In this section, we present three sets of results: the average energy per atom obtained from the raw data of the simulations with thermal kinetic energy of $\frac{3}{2}RT$ added, the heat capacity taken as the derivative of the energy with respect to temperature, and the thermal electronic energy taken from the difference between the equilibrated and low point data.

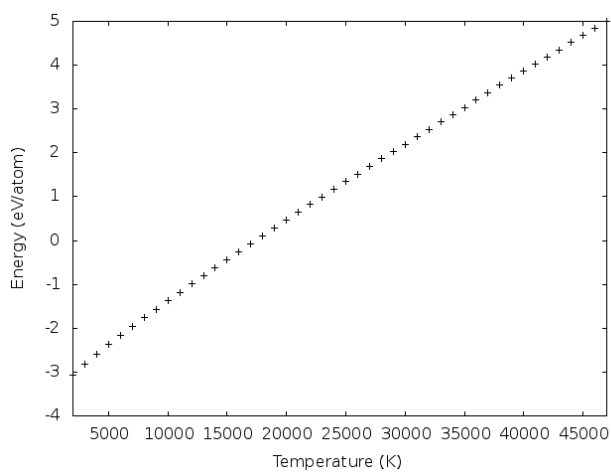

**FIGURE 2.** QMD results for the cold-curve + nuclear components of the energy of Al at 2.7 g/cm³ as a function of temperature, as extracted from the average of the equilibriated and sampled data.

Fig. 2 shows the results obtained by applying the algorithm described above to Al at normal density (2.7 g/cm³) and temperatures ranging from near melt (2,000K) to 47,000K. The energies shown represent the static-lattice contribution plus the thermal nuclear component, $\phi_0(\rho) + F_{nuc}(\rho,T)$. Results are smooth and well-behaved. Predictions for the nuclear component of the constant-volume heat capacity, $C_v^{(nuc)}$, are obtained by differentiating the results from Fig. 2. For this purpose, we fit a Gram polynomial [23,24] (also referred to as piecewise Chebyshev polynomials) to the energy in Fig. 2 using a five-point stencil and then take the derivative of the Gram polynomial, providing a least-squares formula with a parabolic fit. This manner of taking the derivative provides smooth and accurate results with which the original energy can be obtained exactly using the formula for the nondifferentiated

Gram polynomial. Results for $C_v^{(nuc)}$ are shown in Fig. 3. The scatter in the results is likely due to trajectory noise, which is amplified by taking the derivative of the data.

Calculations are compared to three semi-empirical models for $C_v^{(nuc)}$ commonly used for the thermal nuclear components of Eq. 1: the Johnson "generic" model [10], the Chisolm "high-temperature" liquid model [11], and the Kerley CRIS model [12-14]. There are two types of Johnson model considered here, the single-phase model and the multiphase model where the liquid phase has a melting transition just below where the multiphase free energy predicts it to be. The Chisolm and CRIS model use the same parameters as the Johnson multiphase model but with the nuclear model for the liquid changed slightly. These semi-empirical models show that the single-phase model matches the simulation data for temperatures up to 5,000K while the curve of the Chisolm model matches the trend later on. Most notably, the CRIS model appears to be the best fit overall.

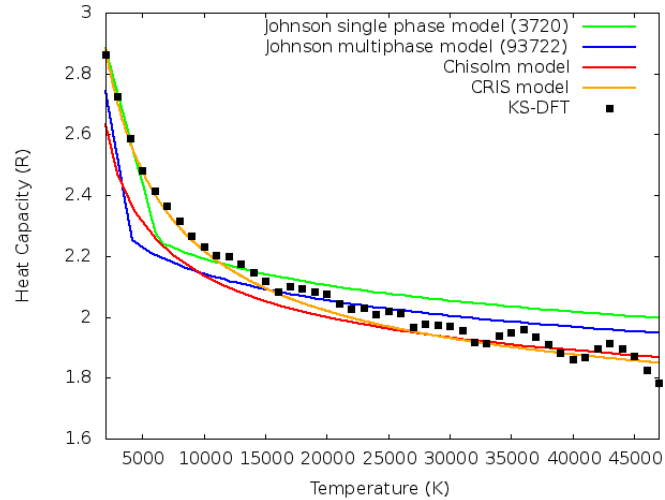

**FIGURE 3.** Results for the constant-volume heat capacity of Al at 2.7 g/cm³ as a function of temperature. QMD results are denoted by black dots; predictions of the Johnson, the Chisolm, and the CRIS models by solid lines.

As a fortunate consequence of this separation of the static-lattice contribution plus the thermal nuclear component from total free energy, we can compute the thermal electronic energy from the difference. These results are labeled 'KS-DFT' in Fig. 4.

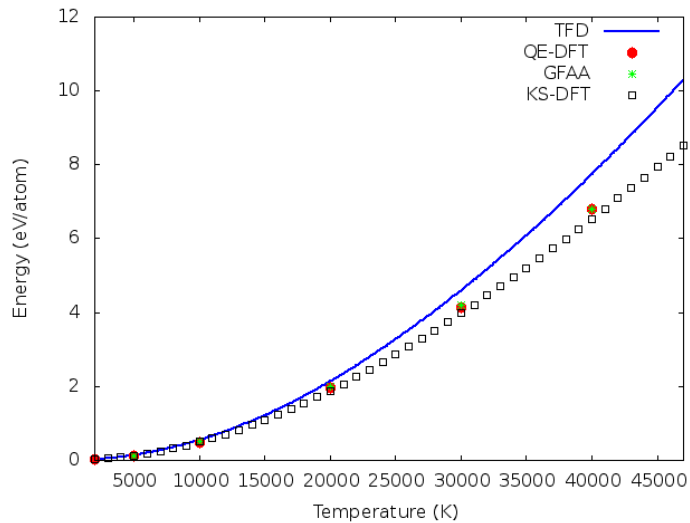

**FIGURE 4.** The thermal electronic energy as a function of temperature in this plot is in units of eV per atom for Al at 2.7 g/cm³. QMD, Quantum-Espresso, and Green's function average atom simulations are marked as points whereas the Thomas-Fermi theory is a line.

In addition to our VASP calculations, Kohn-Sham calculations were performed using the Quantum-Espresso DFT code [25] in which the PBE exchange-correlation functional was employed and a three-valence electron PAW pseudopotential [26] was utilized. The simulations were for a 4 atom FCC primitive cell and converged in a $5^3$ Monkhorst-Pack k-point sampling of the Brillouin zone. The number of bands calculated varied with temperature to ensure a threshold Fermi-Dirac occupation of the highest level of less than $10^{-5}$ for all temperatures. The internal energy was calculated at every temperature and the thermal electron contribution was taken to be the difference between the internal energy and that calculated at 300 K. Though there is some difference between zero and 300 K energies, it is a negligible shift on these curves. The Green's function average atom calculations are Kohn-Sham atoms in jellium systems, but solved via Green's functions instead of through the Kohn-Sham orbitals, which provides a more robust and faster solver [27, 28]. Lastly, the Thomas-Fermi atomic model [27] is included in Fig. 4 which appears to overestimate the electronic energy compared with the three simulations.

## CONCLUSIONS AND FUTURE WORK

The methods presented here demonstrate a tractable way to isolate both the thermal nuclear and thermal electronic components of the heat capacity. Simulations are essential to testing theory regarding the components of heat capacity because experiments are inherently integral measurements and cannot decompose artificial heat capacity components. Future work would be to extend these simulations to even higher temperatures, additional isochores, and other elements. The runs with higher temperatures take longer as more orbitals are required since the Fermi-Dirac distribution becomes wider. Ideally, one would increase the number of processors but this unfortunately has diminishing returns as solving the plane-wave equations in reciprocal space in their matrix form does not scale linearly due to communication issues. Also, when we go to higher temperatures we have to consider the pseudopotential which is limited by the three valence electrons available in VASP's information about aluminum (seen in the 'POTCAR' file). At temperatures high enough, it will be necessary to promote core electrons. Our confidence in the appropriate pseudopotential can be checked by examining the Fermi-Dirac distribution from the output file ('OUTCAR') of the desired temperature and seeing if there appears to be a horizontal gap away from the lowest orbital energy. For our studies, the pseudopotential appeared to be suitable even at the high temperature of 70,000K because the lowest orbital energy (around -5 eV) is occupied. To drive the temperature even higher we can make use of orbital-free DFT for generating a high temperature trajectories and then sample those configurations and run a zero temperature KS-DFT calculation to extract the thermal ionic contribution.

More feasible future work would be to study the liquid regime using aluminum at different densities or doing the same kind of study on another element, like titanium. By having another set of data in the liquid regime we can draw stronger conclusions about the performance of the theoretical models. From this particular aluminum study, it would suggest that the CRIS model performs the most accurately.

Finally, this methodology opens an avenue for examining the accuracy of theories for the thermal electronic contribution – the third term in Eq. 1 – since it is just the difference between the total energy obtained from a conventional QMD simulation and the cold curve and thermal nuclear energy obtained from our algorithm (Fig. 2). This provides a means of assessing approximations in various average atom models (Thomas-Fermi, Purgatorio, etc.).

## ACKNOWLEDGMENTS

The authors gratefully acknowledge Geoffrey Cox, of AWE, for helpful discussions pertaining to the CRIS model and for providing the CRIS predictions for Al. This research used resources provided by the Los Alamos National Laboratory Institutional Computing Program, which is supported by the U.S. Department of Energy National Nuclear Security Administration under Contract No. DE-AC52-06NA25396. This material is based upon work supported by the National Science Foundation Graduate Research Fellowship Program under Grant No. DGE 1322106. Any opinions, findings, and conclusions or recommendations expressed in this material are those of the author(s) and do not necessarily reflect the views of the National Science Foundation.

## REFERENCES

1.  Y. B. Zel'dovich and Y. P. Raiser, *Physics of Shock Waves and High-Temperature Hydrodynamic Phenomena*, edited by W. D. Hayes and R. F. Probstein (Dover, Mineola, NY, 2002), pp. 685-705.
2.  O. L. Anderson, *Equations of State of solids for Geophysics and Ceramic Science* (Oxford, New York, 1995), pp. 3-5.
3.  Z. N. Zharkov and V. A. Kalinin, *Equations of State for Solids at High Pressures and Temperatures*, trans. by A. Tybulewicz (Consultants Bureau, New York, 1971), pp. 1-85.
4.  R. M. More, K. H. Warren, D. A. Young, and G. B. Zimmerman, Phys. Fluids **31**, 3059-3078 (1988).
5.  S. P. Lyon and J. D. Johnson, "Sesame: The Los Alamos National Laboratory Equation of State Database," LA-UR-92-3407, Los Alamos National Laboratory (1992).
6.  G. Kresse and J. Hafner. Phys. Rev. B **47**, 558 (1993).
7.  G. Kresse and J. Hafner. Phys. Rev. B **49**, 14251 (1994).
8.  G. Kresse and J. Furthmüller. Comput. Mat. Sci. **6**, 15 (1996).
9.  G. Kresse and J. Furthmüller. Phys. Rev. B **54**, 11169 (1996).
10. J. D. Johnson, "A generic model for the ionic contribution to the equation of state," *International Journal of High Pressure Research* 6.5 (1991), pp. 277-285.
11. E. Chisolm, S. Crockett, and D. Wallace, "Extending the CCW EOS I: Extending the Cold and Nuclear Contributions to High Compression," LA-UR-03-7344, Los Alamos National Laboratory (2005).
12. G. I. Kerley, J. Chem. Phys. 73, 469-477 (1980)
13. G. I. Kerley, J. Chem. Phys. 73, 478-488 (1980)
14. G. A. Cox and M. A. Christie, J. Phys.: Condens. Matter 27, 405201 (2015).
15. P. Debye,"Zur Theorie der spezifischen Waerme," *Annalen der Physik*, **39** (4), (1912), pp. 789–839.
16. Einstein, "Theorie der Strahlung und die Thoerie der Spezifischen Wärme," *Annalen der Physik*, 4. **22** (1907), pp. 180–190.
17. Petit and P. Dulong, "Recherches sur quelques points importants de la Théorie de la Chaleur". *Annales de Chimie et de Physique*, 10 (1819), pp. 395–413.
18. W. Kohn and L. Sham. "Self-Consistent Equations Including Exchange and Correlation Effects". *Physical Review*. **140** (4A) (1965), pp. A1133–A1138.
19. J. P. Perdew, K Burke, and M. Ernzerhof, Phys. Rev. Lett. 77, 3865 (1996).
20. J. P. Perdew, K. Burke, and M. Ernzerhof, Phys. Rev. Lett., **78** 1396 (1997).
21. "Aluminum." *Periodic Table of Elements: Los Alamos National Laboratory*, periodic.lanl.gov/13.shtml.
22. S. David and J. Steckel, *Density functional theory: a practical introduction*. John Wiley & Sons (2011), pp. 59-60.
23. W. Stephen, and R. L. Pigford, "An approach to numerical differentiation of experimental data," *Industrial & Engineering Chemistry* 52.2 (1960), pp. 185-187.
24. F. Hildebrand, *Introduction to numerical analysis*, Courier Corporation (1987), pp. 350-356.
25. P. Giannozzi, et al., J.Phys.: Condens. Matter, 21, 395502 (2009)
26. A. Dal Corso, Computational Material Science 95, 337 (2014).
27. C. Starrett, "A Green's function quantum average atom model." *High Energy Density Physics* 16 (2015), pp. 18-22.
28. N. Gill and C. Starrett. "Tartarus: A relativistic Green's function quantum average atom code." *High Energy Density Physics* (2017).
29. R. P. Feynman, N. Metropolis, and E. Teller, Phys. Rev. 75, 1561 (1949).